\documentclass[12pt]{article}
\usepackage{fullpage,epsfig,fancyheadings}

\usepackage[psamsfonts]{euscript}

\usepackage{epic}

\usepackage{amstex}

\newcommand{\beq}{\begin{equation}}
\newcommand{\eeq}{\end{equation}}

\font\mybb=msbm10 at 10pt
\def\bb#1{\hbox{\mybb#1}}
\newcommand{\ZZ}{\bb{Z}}
\newcommand{\RR}{\bb{R}}

\begin{document}
\vspace*{-.6in}
\thispagestyle{empty}
{\small
\begin{flushright}
CALT-68-2046\\
hep-th/9604171\\
\end{flushright}}
\baselineskip = 20pt

\vspace{.5in}
{\Large
\begin{center}
Self-Dual Superstring in Six Dimensions\footnote{Work supported in part by the
U.S. Dept. of Energy under Grant No. DE-FG03-92-ER40701.}
\end{center}}
\vspace{.4in}

\begin{center}
John H. Schwarz\\
\emph{California Institute of Technology, Pasadena, CA  91125 USA}
\end{center}
\vspace{1in}

\begin{center}
\textbf{Abstract}
\end{center}
\begin{quotation}
\noindent  A free superstring with chiral $N=2$ supersymmetry in six dimensions
is proposed.  It couples to a two-form gauge field with a self-dual field
strength.  Compactification to four dimensions on a two-torus gives a strongly
coupled N=4  four-dimensional gauge theory with 
$SL(2, \ZZ)$ duality and an infinite tower of dyons. Various authors have suggested
that this string theory should be also the world-volume theory of $M$ theory five-branes.
Accepting this proposal, we find a puzzling factor of two in the application to
black-hole entropy computations.
\end{quotation}
\vfil

\newpage

Evidence has been accumulating for the existence of a new kind of superstring
theory.  This string theory would be non-gravitational, defined in a rigid
six-dimensional background geometry.  Also, it would be self-dual in the sense
that it carries a charge that couples to a two-form $B$ with a self-dual field
strength $H$ $(H = dB = * H$).  Thus, encircling the string with a three-sphere
$S^3$, the magnetic charge $\int_{S^{3}} H$ and the electric charge
$\int_{S^{3}} *H$ are one and the same.\footnote{This possibility was first
noted in Ref.~\cite{duff}.}  This string is also supposed to give a 6D theory
with chiral $N=2$ supersymmetry.  By analogy with the convention in ten
dimensions, I propose to refer to this as a self-dual type IIB string theory.

The first evidence for existence of this theory was put forward by
Witten~\cite{witten}.  He considered the ten-dimensional IIB superstring
theory compactified on $\RR^6 \times K3$, which gives a six-dimensional theory
with  IIB  supersymmetry.  Among the various BPS $p$-brane solitons, he drew
special attention to those that arise from wrapping the self-dual three-brane
on a two-cycle of the $K3$.  This appears as a self-dual string in $\RR^6$ with
IIB supersymmetry and a tension that is proportional to the area of the
two-cycle.  By approaching a point in the moduli space that corresponds to this
area vanishing, the string tension can be made arbitrarily small.  Thus, the
scale of the tension can be decoupled from the Planck scale, and there should
be an effective description of the dynamics of this string in a rigid
space-time background.  The massless modes of the string are easily identified
as a IIB tensor multiplet, which consists of a two-form with self-dual field
strength, five scalars, and two chiral fermions.  It is convenient to describe
states by representations of the little group $SO(4) \approx SU(2) \times
SU(2)$.  In this notation, the tensor multiplet contains $(3,1) + 5 (1,1) +
4(2,1)$.

A second piece of evidence for the existence of this string theory comes from
$M$ theory.  This (still mysterious) theory in eleven dimensions is known to
contain a BPS two-brane and a dual five-brane.  Moreover, it has been noted
that a two-brane can end on a five-brane, where its boundary looks like a
closed string~\cite{strominger}.  This is analogous to the story for
$D$-branes, where (by definition) open strings can end.  Indeed, if one
imagines an open string ending on a Dirichlet 4-brane in the IIA superstring
theory, and recalls that this theory  actually contains a circular eleventh
dimension, it is clear that this configuration corresponds at strong coupling
to a $M$ theory two-brane ending on a five-brane.\footnote{It is less clear to
me what eleven-dimensional interpretation should be assigned to a IIA open
string ending on Dirichlet two-brane.}  This picture has been utilized to argue
that the massless modes of the five-brane world-volume theory should consist of
a IIB tensor multiplet~\cite{becker}.  Moreover, it has been suggested that the complete five-brane world-volume
theory should be a self-dual superstring theory with exactly the properties
discussed above \cite{dijkgraaf}.

Because the superstring charge is self-dual, the two-form field couples to the
string with strength of order unity; therefore, the theory in question is
necessarily strongly coupled.  Without a weak coupling limit analogous to that
of more familiar string theories, one might wonder whether it is possible to
say anything sensible about this theory beyond its massless spectrum.  The key
to a large portion of recent progress has been to focus on BPS saturated states
whose properties at weak coupling often extend unchanged to strong coupling.  This
type of reasoning will play a role in our reasoning later, when we consider
compactifying some of the six dimensions, but in flat infinite six-dimensional
space it is not so useful.  Rather, I would like to suggest that even though
extended classical strings interact strongly, the individual quantum states of
the string do not carry the $B$ charge, and it is consistent to treat them as
free particles.  One way of thinking about this is to imagine compactifying to
four dimensions on $\RR^4 \times T^2$.  Then, as noted in Ref.~\cite{witten},
the $B$ field gives rise to a $U(1)$ gauge field in four dimensions.  Moreover,
winding numbers around the two cycles of the torus correspond to electric and
magnetic charge.  An attractive aspect of this picture is that the $SL(2, \ZZ)$
duality of the resulting N=4  four-dimensional gauge theory acquires a geometric
interpretation in terms of the torus.  This theory, containing both electrically and magnetically
charged states, is certainly strongly coupled.  However, in the
decompactification limit all the charged states become infinitely massive.
Therefore, it seems consistent to consider the $\RR^6$ string as noninteracting.

With the motivation described above, let us now try to construct the desired
free superstring theory.  It is natural for this purpose to work in the space-time
supersymmetric (``Green--Schwarz'') formalism.  When that formalism was first
introduced, it was noted~\cite{green2} that at the classical level the
construction work equally well for $D=10,6,4,3$.  However, manifestly covariant
quantization is extremely difficult, because of the structure of constraints.
Quantization is straightforward in a light-cone gauge, but then it is necessary
to check whether Lorentz invariance is preserved~\cite{green1}.  It was
asserted that this is the case for $D = 10$ only.  Now that a 6D theory is
desired, one wonders what might have been overlooked.

If one does the usual light-cone gauge analysis in six dimensions, the
left-moving or right-moving part of the string spectrum is generated by bosonic
oscillators $\{\alpha_n^i\}$, which are the modes of the transverse spatial
dimensions, and fermionic oscillators $\{S_n^a\}$, which are modes of the
surviving components of the $\theta$ coordinates.  These have the usual
algebras
\begin{eqnarray}
[\alpha_m^i, \alpha_n^j] &=& m\delta_{m + n,0} \delta^{ij}\nonumber\\
\{S_m^a, S_n^b\} &=& \delta_{m+n, 0} \delta^{ab}.
\end{eqnarray}
The index $i$ labels the $(2,2)$ representation of $SU(2) \times SU(2)$ and the
label $a$ labels the $2(2,1)$ representation.  Since there are four of each,
the zero-point energies cancel, and the ground state is massless.  The left-moving
or right-moving component of the massless spectrum 
is given as a representation of the zero-mode Clifford
algebra $\{S_0^a, S_0^b\} = \delta^{ab}$.  This gives a multiplet $2(1,1) +
(2,1)$, which is one-half of a $N=1$ hypermultiplet.  This is a
representation of $N=1$ supersymmetry, but it is not CPT invariant.  This would be a
problem if we were constructing an open string theory, but it need not be one
for a closed-string theory.  The massless IIB spectrum is given by tensoring
this multiplet with itself, and this gives exactly the desired IIB tensor
multiplet, which is CPT invariant.

One might wonder at this point whether this six-dimensional theory is
consistent after all.  To see that it is not, let us consider the first massive level.
This is obtained by applying the raising operators
$\alpha_{-1}^i$ and $S_{-1}^a$ to the ground state and gives the $SU(2) \times
SU(2)$ right-moving content
\begin{equation}
\big( (2,2) + 2(2,1) \big) \times \big( 2(1,1) + (2,1) \big) .
\end{equation}
Since this describes a massive level, these states should combine into
$SO(5) \approx USp(4)$
multiplets.  Such multiplets are non-chiral, which means that they are
invariant under interchange of the two $SU(2)$'s.  This clearly fails here, so
we confirm the claim of Ref.~\cite{green2} that this is not a Lorentz invariant
theory.

The next step is to consider how the preceding construction could be modified
so as to recover 6D Lorentz invariance without changing the massless sector.
The key to answering this question is to recall that the corresponding
light-cone gauge construction in ten dimensions succeeds.  It contains bosonic
coordinates transforming as an 8-vector of $SO(8)$ and fermionic coordinates
transforming as an 8-spinor of $SO(8)$.  With respect to the $SU(2) \times SU(2)$
subgroup considered here, the 8 bosons decompose as $(2,2) + 4(1,1)$ and the 8
fermions decompose as $2(2,1) + 2(1,2)$.  Since we already have oscillators
$\alpha_n^i$ and $S_n^a$ corresponding to the $(2,2)$ and $2(2,1)$, this
suggests introducing new ones corresponding to $4(1,1)$ and $2(1,2)$.  The
simplest way to avoid changing the zero modes is to take these fields to be
antiperiodic on the string.  This means that the oscillators have half-integer
modes.  Thus, we introduce bosonic oscillators $\beta_r^I$ and fermionic
oscillators $T_r^{\dot a}$, where $r\in \ZZ + 1/2$, $I$ labels $4 (1,1)$,
and $\dot a$ labels $2(1,2)$.  I now claim that the resulting spectrum has
six-dimensional Lorentz invariance.  Moreover, the chiral closed-string
spectrum has IIB supersymmetry.  One could check this by representing the
super-Poincar\'e generators in terms of the oscillators and checking the algebra,
as in~\cite{green1}.  This is somewhat tedious, and I have not bothered to do
it.  Instead one can note the following: if one were to compactify the ten-dimensional
IIB superstring on the orbifold $T^4/\ZZ_2$, the twisted sector associated
with any one of the orbifold points would have exactly the content that we have
described, and this is certainly part of a theory with 6D Lorentz
invariance.\footnote{This construction suggests possible generalizations.}
Another easy check is to examine the first few massive levels, and to verify
that they assemble into $USp(4)$ multiplets.  For example, the first massive
level has a right-moving spectrum given by acting with $\beta_{-1/2}^I$ and
$T_{-1/2}^{\dot a}$ on the ground state.  This has $SU(2) \times SU(2)$ content
\begin{equation}
\big( 4(1,1) + 2(1,2) \big) \times \big( 2 (1,1) + (2,1) \big),
\end{equation}
which assembles into two copies of the massive vector supermultiplet, whose
$USp(4)$ content is ${\bf 5} + 3 \cdot {\bf 1} + 2 \cdot {\bf 4}$.

We now seem to have a consistent free theory in six dimensions with the desired
properties.  As we have mentioned, toroidal compactification necessarily turns
it into a strongly interacting theory.  It would be very interesting to find an
efficient way of describing string interactions in that case.

In Ref.~\cite{dijkgraaf} it was proposed that the $M$ theory five-brane is
described by a superstring theory of the type we have described.  The
significant difference is that Ref.~\cite{dijkgraaf} only included $\alpha_n^i$
and $S_n^a$ excitations, and not $\beta_r^I$ and $T_r^{\dot a}$ excitations.
By compactifying $M$ theory on $T^5$ or $T^6$, they computed the BPS 0-brane
spectrum in six and five dimensions.  The five-brane wraps on the torus, and
its modes are described as multi-string configurations.  The BPS spectrum they
found in six dimensions only depends on the self-dual string ground state;
therefore, it is not affected by the addition of $\beta$ and $T$ modes.
However, the five-dimensional spectrum utilizes the entire left-moving
string spectrum.  Specifically, Ref.~\cite{dijkgraaf} 
used the fact that the asymptotic density of states
is characterized by a unitary conformal field theory with $c = 6$.  This
entered into the standard asymptotic degeneracy formula exp $(2\pi
\sqrt{ch/6})$, where $h$ is the level number.  The addition of $\beta$ and $T$
oscillators implies that one should take $c = 12$ instead. Using $c=6$,
Ref.~\cite{dijkgraaf} obtained a $D=5$ black-hole entropy in agreement
with that obtained previously from other considerations.  This result has been
confirmed recently for four-dimensional black holes, as well \cite{klebanov}.
In view of the results presented here, it is puzzling why $c=6$ rather than $c=12$
should give the correct entropy.

I am grateful to R. Dijkgraaf, S. Kachru, H. Nicolai, and H. Ooguri for helpful discussions.

\end{document}